\documentclass[12pt]{article}
\usepackage{epsfig}
\usepackage{graphicx}
\usepackage{amsmath}
\usepackage{amssymb}
\usepackage{slashed}
\usepackage{epstopdf}
\usepackage{multirow}
\usepackage{array}
\usepackage{cite}

\usepackage{xcolor}
\definecolor{lcolor}{rgb}{0.,0.0,0.}
\definecolor{citcolor}{rgb}{0,0.,0.5}
\usepackage[breaklinks,colorlinks,urlcolor=blue,citecolor=blue,linkcolor=blue]{hyperref}
\usepackage{mciteplus}

\numberwithin{equation}{section}

\newcommand{\be}{\begin{equation}}
\newcommand{\ee}{\end{equation}}
\newcommand{\bea}{\begin{eqnarray}}
\newcommand{\eea}{\end{eqnarray}}
\newcommand{\bem}{\begin{multline}}
\newcommand{\eem}{\end{multline}}
\newcommand{\beg}{\begin{gather}}
\newcommand{\eeg}{\end{gather}}

\def\eq#1{{Eq.~(\ref{#1})}}
\def\fig#1{{Fig.~\ref{#1}}}
\def\tabl#1{{Table~\ref{#1}}}
\newcommand{\ben}{\begin{eqnarray*}}
\newcommand{\een}{\end{eqnarray*}}

\newcommand{\chisq}{\chi^2/\mathrm{d.o.f.}}

\begin{document}
\title{{\bf Resummation of double collinear logs in BK evolution versus HERA data \\[0cm] }}
\author{
{\bf Javier L. Albacete }\\  {\it \small CAFPE and Departamento de F\'isica Te\'orica y del Cosmos,  Universidad de Granada}\\ {\it \small 18071 Campus de Fuentenueva, Granada, Spain.} \\ \\[0.0cm] {\texttt{ \small albacete@ugr.es}}}

\date{}
\maketitle

\begin{abstract}

We present a global fit to HERA data on the reduced cross section measured in electron-proton collisions in the region of small Bjorken-$x$: $x\le x_0=10^{-2}$ and moderate to high values of the virtuality $Q^2<Q^2_{max}=650$ GeV$^2$. The main dynamical ingredients in the fits are two recently proposed improved BK equations for the description of the small-$x$ evolution of the dipole scattering amplitude \cite{Beuf:2014uia,Iancu:2015vea}. These two new equations provide an all-order resummation of double collinear logarithms that arise beyond leading logarithmic accuracy. We show that a very good description of data is possible in both cases, provided the parent dipole or smallest dipole prescriptions are employed for the running of the coupling. 
\end{abstract}

\section{Introduction}

The Balitsky-Kovchegov (BK) equation~\cite{Kovchegov:1999yj,Balitsky:1996ub} for the small-$x$ evolution of color dipoles provides the main practical tool for phenomenological studies of non-linear, {\it saturation} effects in available experimental data. The  BK equation corresponds to the large-$N_c$ limit of the B-JIMWLK~\cite{Balitsky:1996ub,Jalilian-Marian:1997gr,Kovner:2000pt,Weigert:2000gi} hierarchy of coupled non-linear evolution equations. These renormalisation group equations are one main pillar of the Color Glass Condensate (CGC) effective theory for the study of QCD scattering process in the high-energy regime. 
 
At leading logarithmic accuracy (LL) the BK equation resums large logarithms $\sim\alpha_s Y$ that arise from the successive emission of small-$x$ gluons. At this degree of accuracy the rapidity variable can be defined as $Y=\ln(x_0/x)$, where $x$ is the Bjorken variable and $x_0$ a reference, initial scale. Additionally, the BK equation also accounts for non-linear, {\it recombination} effects that tame or {\it saturate} the growth of gluon densities with decreasing  $x$. 
Over the last years great efforts have been devoted to systematically improve the dynamical input of the BK and B-JIMWLK equations beyond leading logarithmic accuracy. One big step in this direction was the calculation of running coupling corrections to the BK and B-JIMWLK evolution kernels performed in~\cite{Kovchegov:2006vj,Balitsky:2006wa}. There, the evolution kernel to running coupling accuracy was obtained via an all-order resummation of a subset of next-to-leading corrections (NLO). More recently, full NLO corrections to the BK equation have been calculated  in~\cite{Balitsky:2008zz}. Progress in the refinement of the theoretical formulation of the CGC is not limited to the study of evolution equations. Thus, full or partial next-to-leading corrections to factorisation theorems for particle production in deep inelastic scattering (DIS) and hadronic collisions have been recently presented in~\cite{Balitsky:2010ze,Beuf:2011xd,Chirilli:2011km,Chirilli:2012jd,Chirilli:2013rta,Altinoluk:2014eka}.

Despite the notable progress briefly reported above, it turns out that the perturbative expansion devised in these works shows an unstable behaviour. Phenomenological studies on the effect of partial\cite{Altinoluk:2011qy,Albacete:2012xq} and full\cite{Stasto:2013cha} NLO corrections in single inclusive particle production in proton-proton and proton-nucleus collisions found that those corrections become very large in certain regions of phase space, large transverse momentum and small rapidities, overcoming the LO contribution and even yielding negative values of the cross sections studied in those works. 
These findings on the instability of NLO corrections have been confirmed recently in a first numerical study of the solutions of the  full BK equation at NLO presented in\cite{Lappi:2015fma}. There, it was found that the solutions of the BK equation at NLO accuracy turn negative for a large class of initial conditions relevant for phenomenological applications. 

The lack of convergence and unstable behaviour of the perturbative expansion of high-energy evolution has been identified as due to the presence of large transverse logarithms that arise beyond the leading logarithmic approximation\cite{Lappi:2015fma}. A related problem --large negative contributions arising from transverse logarithms-- was identified after the NLO calculation of BFKL evolution~\cite{Fadin:1995xg,Fadin:1998py,Ciafaloni:1998gs}, the linear limit of the BK equation.

Although their precise derivation is rather technical, the problematic nature of these logarithms can be understood on general physical grounds:
The derivation of the BFKL and BK equation rely on the strong ordering of successively emitted gluons. This separation of time scales paves the way for the factorisation of high-energy evolution: new emissions do not alter the kinematics of previously emitted gluons and can be regarded as independent, i.e their emission factorises. Given that the lifetime of an emitted gluon is $\tau=k^+/2k_{\perp}^2$, this ordering is achieved by imposing strong ordering of the plus light-cone component of successive emissions $q^+\gg k_1^+\gg\dots k_n^+$, provided that their transverse momenta are of the same order $1/r \approx Q_0\approx k_{\perp 1} \approx \dots k_{\perp n}$. Here $1/r$ and $Q_0$ are the characteristic transverse momentum scales of the projectile and target, respectively. In the dipole model of DIS they can be related to the transverse dipole size, $r\sim1/Q$, where $Q$ is the photon virtuality, and the saturation scale of the target $Q_0$. However, the time order condition at the basis of high-energy factorisation is violated by emissions of large transverse momentum, which contribution is parametrically controlled by the the logarithmic extent of the transverse phase space $\rho=\ln(1/rQ_0)$, and  become as large as the leading logarithms when $Y\sim\rho$. Formally, these contributions arise at NLO accuracy as double-logarithmic corrections $\sim \alpha_s Y\rho$ and $\alpha_s \rho^2$, globally referred to as double collinear logs. Clearly, their importance increases for large $Q$-values, i.e in the collinear limit, and their adequate treatment is crucial to match properly with  DGLAP evolution. 

An alternative approach to standard order-by-order calculations consists in rearranging the perturbative expansion via a resummation of large double logarithmic corrections to all orders. Such has been the strategy followed in two recent works: \cite{Beuf:2014uia} and \cite{Iancu:2015vea}.  The net effect of these resummations is to limit the longitudinal and transverse phase space in order to ensure the time ordering condition that lies at the basis of the BK and BFKL equations, thereby restoring the stability of the perturbative series. As we shall see, a direct consequence of this reduced phase space is the further slow down of the evolution speed. 
The work of \cite{Beuf:2014uia} relies in the use of the Mellin space approach and results in the modified BK equation \eq{kc}. This equation explicitly includes a kinematic constraint in the evolution kernel in the form of theta function that reduces the transverse phase space and is non-local in rapidity. We shall refer to it as kinematically corrected BK equation, KC-rcBK.  In turn, the collinearly improved BK equation presented in \cite{Iancu:2015vea}, \eq{dla}, relies in a diagrammatical calculation, was fully derived in coordinate space and is local in rapidity. The locality of the improved BK equation is achieved via an analytic continuation of the physical scattering amplitude outside the physics range $Y\!>\!\rho$. We shall refer to this equation as DLA-rcBK in what follows.     
      
Here we aim at exploring the compatibility of those two collinearly improved BK equations with data on the reduced cross section in e+p collisions at small values of Bjorxen-$x$ measured at HERA by the H1 and ZEUS collaborations\cite{Aaron:2009wt}. 
The role on non-linear corrections in small-$x$ QCD evolution in the interpretation of HERA dada has been thoroughly investigated in previous works\cite{Albacete:2009fh,Albacete:2010sy,Albacete:2012rx,Kuokkanen:2011je,Lappi:2013zma}. The main conclusion extracted from these works is that the BK equation including {\it only} running coupling corrections to the evolution kernel provides a very good description of all available experimental data at small Bjorken-$x$. Further, the precise quantitative information extracted from these fits in the form of parametrizations of the dipole-proton scattering amplitude has become an essential tool to calibrate the physics expectations and to analyse data from other experimental programs, most notably from the proton-proton, proton-nucleus and nucleus collisions performed at RHIC and the LHC (see e.g. \cite{Albacete:2014fwa,Albacete:2012xq}).  

During the completion of this work a similar analysis of HERA data based on the DLA-rcBK equation \eq{dla} has been presented in \cite{Iancu:2015joa}. The fits to HERA data presented in that work rely on a very similar set up to the one followed here, the main methodological differences being the choice of initial conditions and the prescription employed for the running of the strong coupling. Further, the work presented in \cite{Iancu:2015joa} also explores the role of single transverse logarithms $\sim \alpha_s \rho$ in the description of data.  Nonetheless, the results of this work in the form of parametrisations of the dipole-proton scattering amplitude have already been employed in a previous publication~ \cite{Albacete:2015zra} for the calculation of the neutrino-nucleon cross section at ultra-high energies.

This work is structured as follows: In section \ref{colleqs} we discuss the basic elements of the collinearly improved  BK equations proposed in \cite{Beuf:2014uia} and \cite{Iancu:2015vea}. We shall refer to these two equations as kinematically corrected running coupling BK equation (KC-rcBK) and DLA running coupling BK equation (DLA-rcBK) respectively. In section \ref{setup} we review briefly the dipole model of deep-inelastic scattering and describe the numerical set up for the fits, including the description of the free parameters to be fitted to experimental data. The only novelty with respect to the procedure employed in previous works \cite{Albacete:2009fh,Albacete:2010sy,Albacete:2012rx} is the use of a new family of initial conditions for the BK evolution that we shall refer to as {\it pre-scaling} initial conditions. Finally in sections \ref{results} and \ref{conclusions} we present the results of the fits and the conclusions.

\section{Collinearly improved BK equations}
\label{colleqs}
In this section we briefly review the basic elements of the BK equation including both running coupling and double logarithmic corrections. 
Before delving into details, it is important to recall that running coupling corrections and double logarithmic corrections arise from the resummation of two different subsets of next-to-leading terms. Therefore, they can be treated as independent from each other and straightforwardly combined in a single equation by, for instance, replacing the LL kernel in the improved collinear equations by the running coupling kernel. This is exactly the procedure to be followed in this work.

The scattering matrix of a colourless quark-antiquark dipole propagating through the gluon field of a target hadron reads: 
\be
\mathcal{S}({\bf x_0},{\bf x_1};Y)=\frac{1}{N_c}\langle\text{tr}\left\{ U({\bf x_0})U^{\dagger}({\bf x_1})\right\}\rangle_Y \equiv \mathcal{S}_{{\bf 01};Y} 
\label{S}
\ee
where ${\bf x_{0,(1)}}$ are the transverse coordinates of the quark and antiquark respectively. Under the eikonal approximation the propagation of each individiual right-moving parton through the hadron target is accounted for time ordered Wilson lines:
\be
U({\bf x})= \text{P}\exp\left[ig\!\int\! dx^- \!A^+(x^-,\bf x)\right]\,,
\ee   
where $A^+$ denotes the gluon field of the target hadron. The average in \eq{S} is performed over the target gluon field configurations at a given rapidity $Y\equiv \ln(x_0/x)$. 
We have used the last line of \eq{S} to introduce the reduced notation for the average of the dipole scattering matrix that we shall employ hereafter.  The BK equation provides the rapidity evolution of the dipole scattering matrix. It reads
\be
\frac{\partial \mathcal{S}_{{\bf 0 1};Y}}{\partial Y}= \int \frac{d^2{\bf x_2}}{2\pi}\, \mathcal{M}_{\bf 012}\left[\mathcal{S}_{{\bf 0 2};Y}\,\mathcal{S}_{{\bf 1 2};Y} - \mathcal{S}_{{\bf 0 1};Y}   \right]\,,
\label{bk}
\ee
where $\mathcal{M}_{\bf 012}$ is the evolution kernel. The BK equation at leading logarithmic accuracy resums large logarithmic corrections of the form $\sim (\alpha_s Y)^n$ to all orders. It accounts for the multiple emission of soft gluons and also for the possibility of gluon recombination via the non-linear term in the right hand-side of \eq{bk}.
The BK equation assumes that the average of the product  of two dipole scattering matrices over the field configurations of the target factorizes:  $\langle S_{xz} S_{zy} \rangle_Y\rightarrow \mathcal{S}_{{\bf xz};Y}  S_{\bf{zy};Y}$, i.e. it is derived in the mean field limit. We shall rely on the translational invariant approximation, equivalent to the assumption that de dipole amplitude only depends on its relative transverse size. We tus introduces the variables $r_0=|\bf{x_0-x_1}|$, $r_1=|\bf{x_1-x_2}|$ and $r_2=|\bf{x_0-x_2}|$.  

The calculation of running coupling corrections to the original LL kernel was performed in \cite{Kovchegov:2006vj,Balitsky:2006wa}. There, two different prescriptions were proposed  
for the kernel of the BK equation at running coupling accuracy. It was shown in \cite{Albacete:2007yr} that Balitsky's prescription~\cite{Balitsky:2006wa} minimizes the role of higher-order, {\it conformal} corrections, suggesting that it may be better suited for phenomenological applications. In particular, Balitsky's prescription \eq{bker} yields a slower evolution speed than other possible schemes explored in the literature, like the Kovchegov-Weigert~\cite{Kovchegov:2006vj} one or the {\it smallest dipole size} prescription, where the scale for the running coupling is set by the smallest of the transverse dipole sizes involved in one evolution step: $r_0$, parent dipole and $r_{1,2}$, daughter dipoles. This feature was crucial for the very good description of previous AAMQS fits to HERA data. However, it remains to be clarified which running coupling scheme is theoretically better motivated once other dynamical effects, like the double logarithmic corrections discussed here, are also incorporated to the BK equation. Under Balitsky's prescription the running coupling kernel reads
\be
\mathcal{M}^{\text{Bal}}_{\bf 012}=\frac{\alpha_s(r_0^2)\,N_c}{\pi}\,\left[\frac{r_0^2}{r_1^2r_2^2}+\frac{1}{r_1^2}\left(\frac{\alpha_s(r_1^2)}{\alpha_s(r_2^2)} -1\right) +\frac{1}{r_2^2}\left(\frac{\alpha_s(r_2^2)}{\alpha_s(r_1^2)} -1\right)  \right] \quad :: \quad \text{Bal}\,.
\label{bker}
\ee 
Here we will consider  two other possibilities for the running coupling kernel, namely the {\it parent dipole} prescription:
\be
\mathcal{M}^{\text{pd}}_{\bf 012}=\frac{\alpha_s(r_0^2)\,N_c}{\pi}\, \frac{r_0^2}{r_1^2r_2^2} \quad :: \quad \text{PD}\,,
\label{pdker}
\ee 
and the {\it smallest dipole} prescription:
\be
\mathcal{M}^{\text{pd}}_{\bf 012}=\frac{\alpha_s(r^2_{\text{min}})\,N_c}{\pi}\, \frac{r_0^2}{r_1^2r_2^2}\,\quad \text{with} \quad r_{min}\equiv \text{min}\{ r_0, r_1, r_2\}\quad :: \quad \text{SD}\,.
\label{sdker}
\ee 
Although the ansatzs \eq{pdker} and \eq{sdker} do not follow from any strict diagrammatical calculation, it was shown in \cite{Albacete:2007yr} that the parent dipole description leads to solutions very similar to those obtained under the prescription derived by Kovchegov and Weigert \cite{Kovchegov:2006vj}. In particular, the {\it parent dipole} prescription leads to significantly faster evolution speed that Balitsky's prescription. We shall use it here, rather than the full Kovchegov-Weigert kernel, due to its relative simplicity in the numerical evaluation. In turn, the smallest dipole prescription is motivated by the expectation that the scale for the running coupling should be given by the hardest momentum scale in the process. 

\subsection{Kinematically corrected BK equation (KC-rcBK)} 

A kinematically improved version of the BK equation consistent at high, but finite, energies has been proposed in \cite{Beuf:2014uia}. It reads:

\be
\frac{\partial \mathcal{S}_{{\bf 0 1};Y}}{\partial Y}= \int \frac{d^2{\bf x_2}}{2\pi}\, \mathcal{M}_{\bf 0 1 2}\,\Theta(Y-\Delta_{\bf 012})\left[\mathcal{S}_{{\bf 0 2};Y-\Delta_{\bf 0 1 2}}\,\mathcal{S}_{{\bf 1 2};Y-\Delta_{\bf 0 1 2}} - \mathcal{S}_{{\bf 0 1};Y}   \right]\,.
\label{kc}
\ee
The reduction of phase space for the evolution is made explicit by the presence of the theta function in \eq{bk}, which constrains gluon emission to some bounded domain in the ${\bf x_2}$-plane. This is in clear contrast with the LL BK equation or the BK equation with only running coupling corrections, where emission of gluons in all the transverse plane are allowed.  The parameter that controls the extent of kinematic corrections is given by: 
\be
\Delta_{\bf 012}= \text{max}\left\{ 0, \ln\left(\frac{l_{\bf 012}^2}{r_0^2} \right) \right\}\,\quad\text{with} \quad l_{\bf 012}=\text{min}\left\{ r_1,r_2 \right\}.
\label{delta}
\ee
There is actually some freedom in the definition of $l_{\bf 0 1 2}$ introduced in \eq{delta}. It should satisfy $l_{\bf 0 1 2}\approx r_1 \approx r_2$ in the regime $r_0\ll r_1 \approx r_2$.  This freedom in the choice of $\Delta_{\bf 0 1 2}$ should be considered as a resummation scheme ambiguity associated with the kinematic constraint. In what follows we shall adopt the definition presented in the r.h.s of \eq{delta}; we have checked that changes in this prescription do not alter significantly the results of the fits. Other important feature of the kinematically corrected BK equation \eq{delta} is that the scattering amplitude of the two newly created dipoles after one evolution step are evaluated at a delayed rapidity $ Y-\Delta_{\bf 0 1 2}$. This rapidity {\it veto} has previously discussed in the literature as a main part of the NLO or energy-momentum corrections to the BK equation~\cite{Gotsman:2004xb,Kuokkanen:2011je}. As we shall see in Section \ref{results} both effects tend to decrease the evolution speed, i.e. to slow down the growth of the saturation scale with decreasing Bjorken-$x$, but they also modify more exclusive features of the unintegrated gluon distribution. Another subtle point that arises in the definition of the high-energy factorisation scheme once the kinematic corrections are taken into account is related to the very definition of the evolution variable, i.e. the rapidity variable. We will not delve into the details here (see the extended discussion in \cite{Beuf:2014uia}). It is however important to recall that the usual choice $Y=\ln ( k_f^+/ k_0^+)=\ln(x_0/x)$  can be modified by finite corrections, leading to different evolution equations at NLO accuracy and beyond. This freedom to choose the evolution variable is related to freedom to choose a reference energy scale in the BFKL formalism. In practice we will deal with this ambiguity through the use of pre-asymptotic initial conditions, see \ref{setup}. The introduction of an arbitatry rapidity shift $\Delta Y_0$ as another free fit parameter copes effectively with a possible redefinition of the rapidity variable. 

\subsection{BK equation at Double Logarithmic Accuracy (DLA-rcBK)}
A BK equation that resums double collinear logs to all orders has been recently proposed in \cite{Iancu:2015vea}. It reads: 
\be
\frac{\partial \mathcal{\tilde{S}}_{{\bf 0 1};Y}}{\partial Y}= \int \frac{d^2{\bf x_2}}{2\pi}\, \mathcal{M}_{\bf 012} \, \mathcal{K}^{\text{DLA}}_{\bf 012} \left[\mathcal{\tilde{S}}_{{\bf 0 1};Y}\,\mathcal{\tilde{S}}_{{\bf 1 2};Y} - \mathcal{\tilde{S}}_{{\bf 0 1};Y}   \right]\,.
\label{dla}
\ee
It should be noted that the object evolved in \eq{dla}, is not the physical dipole scattering matrix $\mathcal{S}_{{\bf 0 1};Y}$, but rather a related function $\mathcal{\tilde{S}}_{{\bf 0 1};Y}$. Both functions are expected to coincide in the physical range $Y>\rho$. For general positive values of $Y$ and $\rho$ the relation between these two quantities is given by
\be
\mathcal{\tilde{A}}(Y,\rho)\equiv \int_0^\rho d\rho_1 \, \tilde{f}(Y,\rho-\rho_1) \,\mathcal{A}(0,\rho_1)\, 
\label{a}
\ee
with
\be
\tilde{f}(Y=0,\rho)=\delta(\rho)-\sqrt{\bar{\alpha}_s}\,\text{J}_1(2\sqrt{\bar{\alpha}_s \rho^2})\,,
\ee
where $\text{J}_1$ is the Bessel function of the first kind. The function $\mathcal{A}(Y,\rho)$ in \eq{a} is defined as $(1-\mathcal{S}_{{\bf x y};Y})\equiv r^2Q_0^2 \mathcal{A}_{{\bf x y};Y} $, with $\rho\equiv \ln(1/r^2Q_0^2)$ and $Q_0$ some initial scale (its precise definition is given in section \ref{setup}). The introduction of the auxiliary function $\mathcal{\tilde{S}}_{{\bf 0 1};Y}$ obeys the purpose of obtaining a collinearly improved equation that, contrary to \eq{kc}, is local in rapidity. This implies performing an analytic continuation of the dipole scattering matrix outside the physical range $Y>\rho$. \eq{dla} exhibits two important features. First, it is local in rapidity. Second, the resummation of double logarithmic corrections results in just a modification of the evolution kernel by the factor 
\be
\mathcal{K}^{\text{DLA}}_{\bf 012} = \frac{\text{J}_1(2\sqrt{\bar{\alpha}_s\rho'^2})}{\sqrt{\bar{\alpha}_s}\rho'^2} \; \quad\text{with}\quad \rho'=\sqrt{\ln\left(r_1^2/r_0^2 \right)\,\ln\left(r_2^2/r_0^2 \right)}.
\label{kdla}
\ee
It is important to note that the resummation affects both the evolution kernel and the initial conditions for the evolution. The latter, $\mathcal{\tilde{A}}$, are obtained applying \eq{a} to the initial conditions defined in Eqs. (\ref{ic}) - (\ref{ic2}) in Section \ref{setup}.  
Finally, the derivation of \cite{Iancu:2015vea} treats the strong coupling as a fixed parameter. In order to be consistent with the degree of accuracy of the evolution kernel(s) \eq{bker} and \eq{pdker} we shall let the coupling run in Eqns. (\ref{a})-(\ref{kdla}) at the scale $\mu=2/r$ or $\mu=2/\text{min}(r,r_1,r_2)$. As we shall discuss in section \ref{results}, it turns out that the results of the fits are very little sensitive to this choice.

\section{Set up}
\label{setup}
\subsection{Dipole model of DIS}
In this section we briefly review the main ingredients needed for the
calculation of the inclusive and longitudinal DIS structure functions, which was extensively discussed previous papers. see e.g. \cite{Albacete:2010sy,Albacete:2009fh}. Neglecting the contribution from $Z$ boson exchange, the reduced cross section can be expressed in terms of the inclusive, $F_2$, and longitudinal, $F_L$, structure functions:
\begin{equation}
\sigma_{r}(y,x,Q^2)=F_2(x,Q^2)-\frac{y^2}{1+(1-y)^2}F_L(x,Q^2),
\label{rcs}
\end{equation}
where $y=Q^2/(s\,x)$ is the inelasticity variable and $\sqrt{s}$  the center of mass collision energy.
In turn, at $x\ll 1$, the inclusive and longitudinal structure functions can be expressed as
\be
F_2(x,Q^2)=\frac{Q^2}{4\,\pi^2\alpha_{em}}\left(\sigma_T+\sigma_L\right)\,,\quad\quad F_L(x,Q^2)=\frac{Q^2}{4\,\pi^2\alpha_{em}}\,\sigma_L\,.
\label{f2l}
\ee
Here $\sigma_{T,L}$
stands for the virtual photon-proton cross section for transverse ($T$)
and longitudinal ($L$) polarization of the virtual photon. In the dipole model, valid at high energies or small $x$, one writes  \cite{Nikolaev:1990ja,Mueller:1989st}:
\begin{equation}
  \sigma_{T,L}(x,Q^2)=\sum_f\int_0^1 dz\int  \,d^2{\bf r}\,\vert
  \Psi_{T,L}^f(e_f,m_f,z,Q^2,{\bf r})\vert^2\,
  \sigma^{q\bar{q}}({\bf r},x)\,,\,,
\label{dm1}
\end{equation}
where $\Psi_{T,L}^f$ is the light-cone wave function
for a virtual photon to fluctuate into a quark-antiquark dipole of quark flavor $f$. Note that $\Psi_{T,L}^f$ only depends on the quark flavor $f$ through the quark mass $m_f$, and electric charge $e_f$ (see e.g.  \cite{Golec-Biernat:1998js} for explicit expressions to lowest
order in $\alpha_{em}$). According to the optical theorem, the dipole cross section $\sigma^{q\bar{q}}({\bf r},x)$ is given by the integral over impact parameter of the (imaginary part of) dipole-hadron scattering amplitude. Further, under the approximation that the dipole scattering amplitude is independent on the impact parameter of the collision one gets
\be
\sigma^{q\bar{q}}(r,x)=2\int d^2b \,{\cal N}(x,r,b) = \sigma_0  \,{\cal N}(x,r)\, 
\ee
where $\sigma_0$ has the meaning of (half) the average transverse area of the quark distribution in the transverse plane and will be one of the free parameters in the fit. 
Following \cite{Albacete:2010sy} we have consider a variable number of active flavours up to $N_f$=5, with a current quark mass $m_{u,d,s,c,b}=0.05$, $0.05$, 0.14, 1.3 and 4.5 GeV respectively. It turns out that the fits results are very little sensitive to modifications of the particular choice made for the current mass of light flavours.

\subsection{Pre-asymptotic initial conditions}
In order to solve the BK equations under the different evolution schemes discussed in section \ref{colleqs} one has first to specify the initial conditions for the evolution at the highest value of Bjorken-$x$ included in the data fitting set, $x_0=0.01$.
The free parameters in the AAMQS fits mainly correspond to the free parameters in the initial conditions for the
evolution, as follows:
\begin{equation}
\mathcal{N}(r, Y=0)=
1-\exp\left[-\frac{\left(r^2\,Q_{0}^2\right)^{\gamma}}{4}\,
  \ln\left(\frac{1}{\Lambda_{QCD}\,r}+e\right)\right]\ ,
\label{ic}
\end{equation}
$Q_{0}$ and $\gamma$ are two free parameters to be fitted to experimental data. $\gamma$ is a dimensionless parameter
that controls the steepness of the unintegrated gluon distribution at moderate and high transverse momenta. In the original AAMSQ fits, $Q_0$ plays the role of the initial saturation scale at the highest value of Bjorken-$x$ included in the fitting data set, $x_0=10^{-2}$ or, equivalently at $Y=0$ with $Y=\ln(x_0/x)$.
However, it turns out that it is not possible to obtain good fits for the collinearly improved equations discussed here using the two-parameter family of initial conditions provided by \eq{ic}. 
In order to allow a larger  freedom for the functional forms of the initial conditions we shall introduce a new family of {\it pre-asymptotic} initial conditions as follows:
We first solve the BK equation using \eq{ic} as the initial conditions for the evolution and take the dipole scattering amplitude at $x=x_0$ as the solution of the BK equation at rapidity $\Delta Y_0$:
\begin{equation}
\mathcal{N}(r,x_0=0.01)=\mathcal{N}(r,\Delta Y_0)\,.  
\label{ic2}
\end{equation}
That is,  we allow the evolution to run for some rapidity interval $\Delta Y_0$ before comparing to experimental data. In other words, we use the evolution itself to generate the initial conditions for further evolution at $Y>\Delta Y_0$ or, equivalently for $x<x_0$. Thus, the dipole scattering amplitude in the physical region $x<x_0$ is given by
\begin{equation}
\mathcal{N}(r,x\le x_0)=\mathcal{N}(r,\Delta Y_0+\ln(x_0/x))\,,  
\label{ic3}
\end{equation}
where $\Delta Y_0$ quantifies the amount of pre-evolution allowed before comparing to experimental data and is another fitting parameter. Several comments are in order:

First, this set up should just be regarded as a mathematical procedure to generate a family of initial conditions which are a solution of the BK equation itself. Therefore, for values $\Delta Y_0\!>0\!$ neither the fitting parameters $Q_0$ and $\gamma$ nor the solutions of the BK equation for $Y\!<\!\Delta Y_0$ afford a clear, straightforward physical interpretation. Thus the scale $Q_0$ in \eq{ic} is just an auxiliary scale to generate the physical initial conditions at $x_0=10^{-2}$. The physically meaningful object is the dipole scattering amplitude for $x\le x_0$ or $Y\ge\Delta Y_0$ as defined in \eq{ic2} and \eq{ic3}. Analogously,  the physically meaningful saturation scale is defined via the condition
\be
\mathcal{N}(r=1/Q_s(x),x)=0.5
\label{qsx}
\ee
for $x\le x_0$. 

Second, similar initial conditions have been previously used in the literature in the limiting case of $\Delta Y_0\to \infty$. These are referred to as {\it scaling} initial conditions in \cite{Kuokkanen:2011je,Albacete:2009fh}. It is well known that the non-linear character of the BK equation leads to scaling solutions, i.e independent of the initial conditions at asymptotically large rapidities. In that limit, the solutions depend only on a single variable, the scaling variable $\tau\equiv r Q_s(x)$, hence effectively becoming a become a one-parameter family of possible initial conditions.  This property of the BK equation has been studied in depth in relation to the observed phenomenon of geometric scaling featured by experimental data on e+p collisions from HERA. Here we take an intermediate approach where we allow for a finite amount of evolution $\Delta Y_0$ before comparing to data. In this intermediate limit the initial conditions generated at $x=x_0$ according to \eq{ic2} are still sensitive to the initial fitting parameters $Q_0$ and $\gamma$.

Finally, the pre-evolution interval $\Delta Y_0$ helps the convergence of the solution of the DLA equation to its regime of physical applicability, defined by the condition $Y>\rho=\ln(1/r^2Q_0^2)$ and,  in a related way, this approach copes with the uncertainty in the definition of the evolution variable, i.e the rapidity  $Y$, in the evolution scheme including kinematic corrections.

\subsection{Regularisation of the infrared dynamics}
The solution of the BK equation in either of the evolution schemes described above implies the evaluation of the strong coupling at 
arbitrarily large values of the dipole size
(small gluon momentum), and a regularization prescription to avoid the
Landau pole becomes necessary. Following previous AAMQS works, for small dipole sizes $r<r_{fr}$
we shall evaluate the running coupling according to the usual one-loop QCD expression:
\begin{equation}
\alpha_s(r^2)=\frac{12\pi}{\left(11N_c-2N_f\right)\,\ln\left
    (\frac{4\,C^2}{r^2 \Lambda_{QCD}^2}\right)}\,.
\label{alpha}
\end{equation}
The number of active flavors $N_f$ in \eq{alpha} should is set to the number of quark flavors lighter than the momentum scale associated with the scale $r^2$ at which the coupling is evaluated $\mu^2= 4C^2/r^2$. The setup of this variable flavor scheme is completed by matching the branches of the coupling with adjacent $N_f$ at the scale corresponding to the quark masses $r^2_*= 4C^2/m_f^2$. For the 1-loop accuracy at which the coupling \eq{alpha} is evaluated, the matching condition is simply given by
\be
\alpha_{s,N_f-1}(r_*)=\alpha_{s,N_f}(r_*)\,.
\ee
With only three active flavours one gets $\Lambda_{QCD}=0.241$ GeV, such that $\alpha_s(M_{Z})=0.1176$, with
$M_{Z}$ the mass of the $Z$ boson. In turn, for larger sizes, $r>r_{fr}$, we freeze the
coupling to the fixed value $\alpha_s(r_{fr})\equiv \alpha_{fr}$. We shall use two different values $\alpha_{fr}=0.7$ and 1. 
The fudge factor $C$ under the
logarithm in \eq{alpha} will be one of the free parameters in the
fit. It reflects the uncertainty in the Fourier transform from
momentum space, where the original calculation of $\alpha_sN_f$
corrections was performed, to coordinate space. We shall extend the fits up to values of the photon virtuality $Q^2\sim 650$ GeV$^2$. 


\section{Results}
\label{results}

\begin{table}
  \centering
  \begin{tabular}{| c | >{\centering}p{1.7cm} | c | c | c | c | c | c | c |}
     \multicolumn{9}{ c }{rcBK} \tabularnewline 
     \hline
   $Q_{max}^2$ $(\mathrm{GeV}^2)$  & Evolution scheme & $\alpha_{fr}$ &  $Q_0^2 (\mathrm{GeV}^2)$ &  $\Delta Y_0$ & $\sigma_0$ (mb) & $ \gamma $ & $C$ & $\chisq$  \\ \hline\hline
    
    \multirow{4}{*}{50} & Bal        &  0.7 &  0.210   & 0 &    24.03      & 1.141   &    1.437  & 0.881       \tabularnewline \cline{2-9} 
                                  &  Bal      &  0.7 &  0.182    &    0   &   24.06  & 1.071   &    1.36 & 1.119      \tabularnewline \cline{2-9} 
                                   & Bal       & 1 &  0.213   & 0   &   24.02   & 1.147   &    1.147  & 0.875                         \tabularnewline \cline{2-9} 
                                   & Bal	& 1 &  0.197 &  0   &   25.00   & 1.101   &    1.39  &0.924                         \tabularnewline \cline{1-9} 

    \multirow{4}{*}{650} & Bal   & 0.7  &   0.216   & 0  &   24.16   & 1.160   &    1.461  & 0.928                           \tabularnewline \cline{2-9} 
                                   & Bal       &  0.7 &   0.209   &  0   &   23.14   & 1.124   &    1.250  & 1.08                           \tabularnewline \cline{2-9} 
                              & Bal       &  1 &  0.218   &  0   &   23.94  & 1.158   &    1.232  &  0.869                          \tabularnewline \cline{2-9} 
                                   & Bal	&  1  &  0.207  & 0   &   23.14   & 1.122   &    1.12  &1.119                        \tabularnewline \cline{1-9}

  \end{tabular}
\caption{Fit parameters for rcBK evolution with the running coupling kernel evaluated according to Balitsky's prescription \eq{bker}.}. 
\label{t1}

 
 \begin{tabular}{| c | >{\centering}p{1.7cm} | c | c | c | c | c | c | c |}
     \multicolumn{9}{ c }{DLA-rcBK} \tabularnewline 
     \hline
   $Q_{max}^2$ $(\mathrm{GeV}^2)$  & Evolution scheme & $\alpha_{fr}$ &  $Q_0^2 (\mathrm{GeV}^2)$  &  $\Delta Y_0$ & $\sigma_0$ (mb) & $ \gamma $ & $C$ & $\chisq$  \\ \hline\hline
    
    \multirow{4}{*}{50} & PD    &  0.7 &  4.72$\cdot 10^{-3}$   & 9.05 &    22.68  & 0.938  & 3.662  & 0.996   \tabularnewline \cline{2-9} 
                                &  Bal   &  0.7 &  3.16$\cdot 10^{-2}$     &  3.21   &   22.79  & 0.810  & 0.566 &    1.531       \tabularnewline \cline{2-9} 
                               & PD   & 1   &  2.21$\cdot 10^{-2}$       & 6.60 &   21.93   & 1.044   &    3.108  &  1.089       \tabularnewline \cline{1-9} 
  
    \multirow{4}{*}{650} & PD   & 0.7  &  2.70$\cdot 10^{-2}$    & 7.09  &   22.54   & 1.1469   &   3.78  & 1.157  \tabularnewline \cline{2-9} 
                              & PD   &  1 &   2.43$\cdot 10^{-2}$   &  7.13   &   22.05  & 1.127   &    3.44  &  1.13      \tabularnewline \cline{2-9} 
	                      & SD   &  0.7 &   5.38$\cdot 10^{-2}$  &  4.19   &   22.91  & 1.166   &    2.69  &  1.093      \tabularnewline \cline{1-9}  
 \end{tabular}
\caption{Fit parameters for DLA-rcBK evolution with the running coupling kernel evaluated according to the Bal, PD and SD schemes.}. 
\label{t2}


 \begin{tabular}{| c | >{\centering}p{1.7cm} | c | c | c | c | c | c | c | c |}
     \multicolumn{9}{ c }{KC-rcBK} \tabularnewline 
     \hline
   $Q_{max}^2$ $(\mathrm{GeV}^2)$  & Evolution scheme & $\alpha_{fr}$ &  $Q_0^2 (\mathrm{GeV}^2)$  &  $\Delta Y_0$ & $\sigma_0$ (mb) & $ \gamma $ & $C$ & $\chisq$  \\ \hline\hline
    
    \multirow{4}{*}{50} & PD    &  0.7 &  4.65$\cdot 10^{-2}$  & 4.98 &    22.33  & 1.001  & 3.662  & 1.081   \tabularnewline \cline{2-9} 
                               & Bal  & 0.7 &  1.66$\cdot 10^{-2}$   &  5.53  &   23.75   & 0.845   &    0.869  &1.332          \tabularnewline \cline{2-9}  
                               & PD   & 1   &  3.35$\cdot 10^{-2}$   & 6.60 &   22.23   & 1.032   &    3.806  &  1.108       \tabularnewline \cline{2-9} 
                               &  Bal   &  1 &  2.07$\cdot 10^{-2}$   &  6.15   &   22.85  & 0.8542  & 0.675 &    1.246   \tabularnewline \cline{1-9} 
 
    \multirow{4}{*}{650} & PD   & 0.7  &  5.22$\cdot 10^{-2}$  & 5.03  &   22.79   & 1.058   &   3.67  & 1.372  \tabularnewline \cline{2-9} 
                                 & PD   &  1 &   0.33$\cdot 10^{-2}$   &  6.14   &   22.19  & 1.040  &    3.74  &  1.280     \tabularnewline \cline{2-9} 
                               & SD   &  0.7 &   0.101   &  3.24   &   21.55  & 1.101  &    2.52  &  1.21     \tabularnewline \cline{1-9} 
 \end{tabular}

 \caption{Fit parameters for KC-rcBK evolution with the running coupling kernel evaluated according to the Bal, PD and SD schemes.}. 
\label{t3}
\end{table}

\begin{figure}[htb]
\begin{center}
\includegraphics[width=13cm]{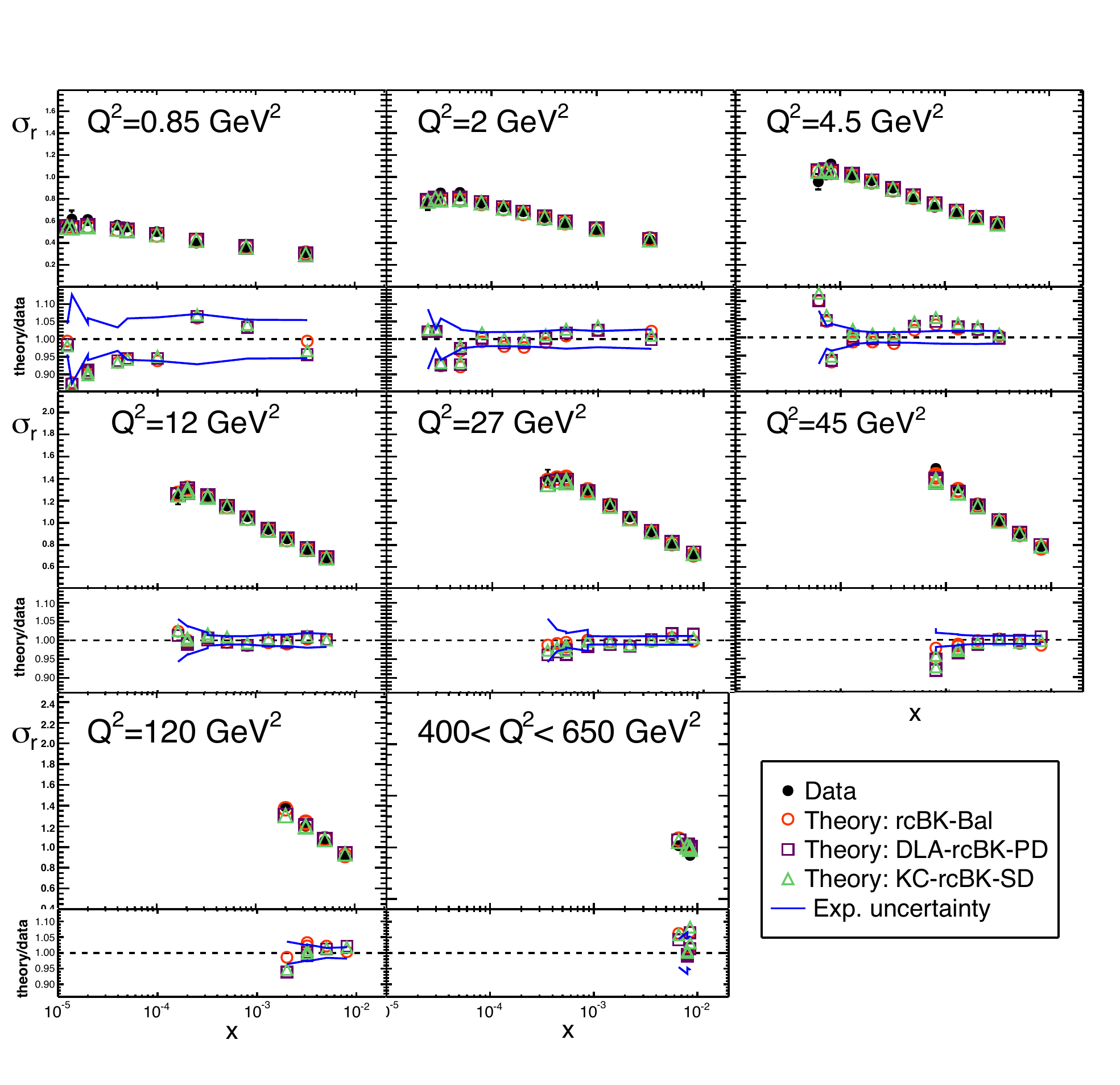}
\end{center}
\vspace*{-0.5cm}
\caption[a]{Comparison between data on the reduced cross section (full circles) and the fit results for three different evolution schemes: rcBK-Bal (open circles), DLA-rcBK-PD (open squares) and KC-rcBK-SD (open triangles). The three fit results correspond to $\alpha_{fr}=0.7$. The lower panels in each plot show the ratio theory/data together with the experimental uncertainty. }
\label{fits}
\end{figure}

With the set up described in the previous section it is possible to obtain good fits to data on the reduced cross sections with either of the evolution schemes rcBK, KC-BK and DLA-BK.  The corresponding fit parameters and the $\chi^2/d.o.f$ are shown in \tabl{t1}-\ref{t3}, whereas a comparison between the fit results and experimental data for some selected $Q^2$ bins is shown in \fig{fits}. 

\begin{figure}[htb]
\begin{center}
\includegraphics[width=8.5cm]{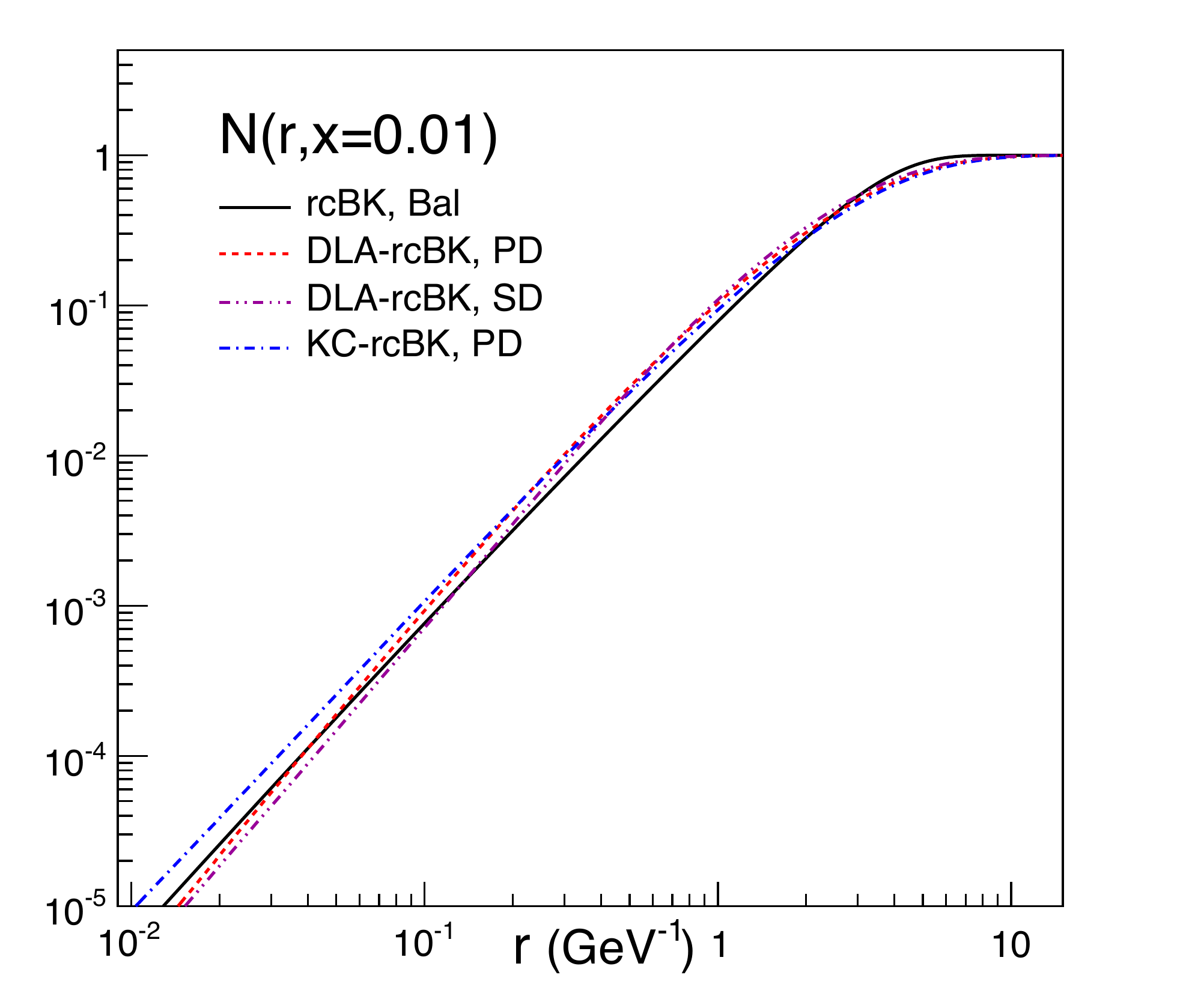}
\includegraphics[width=8.25cm]{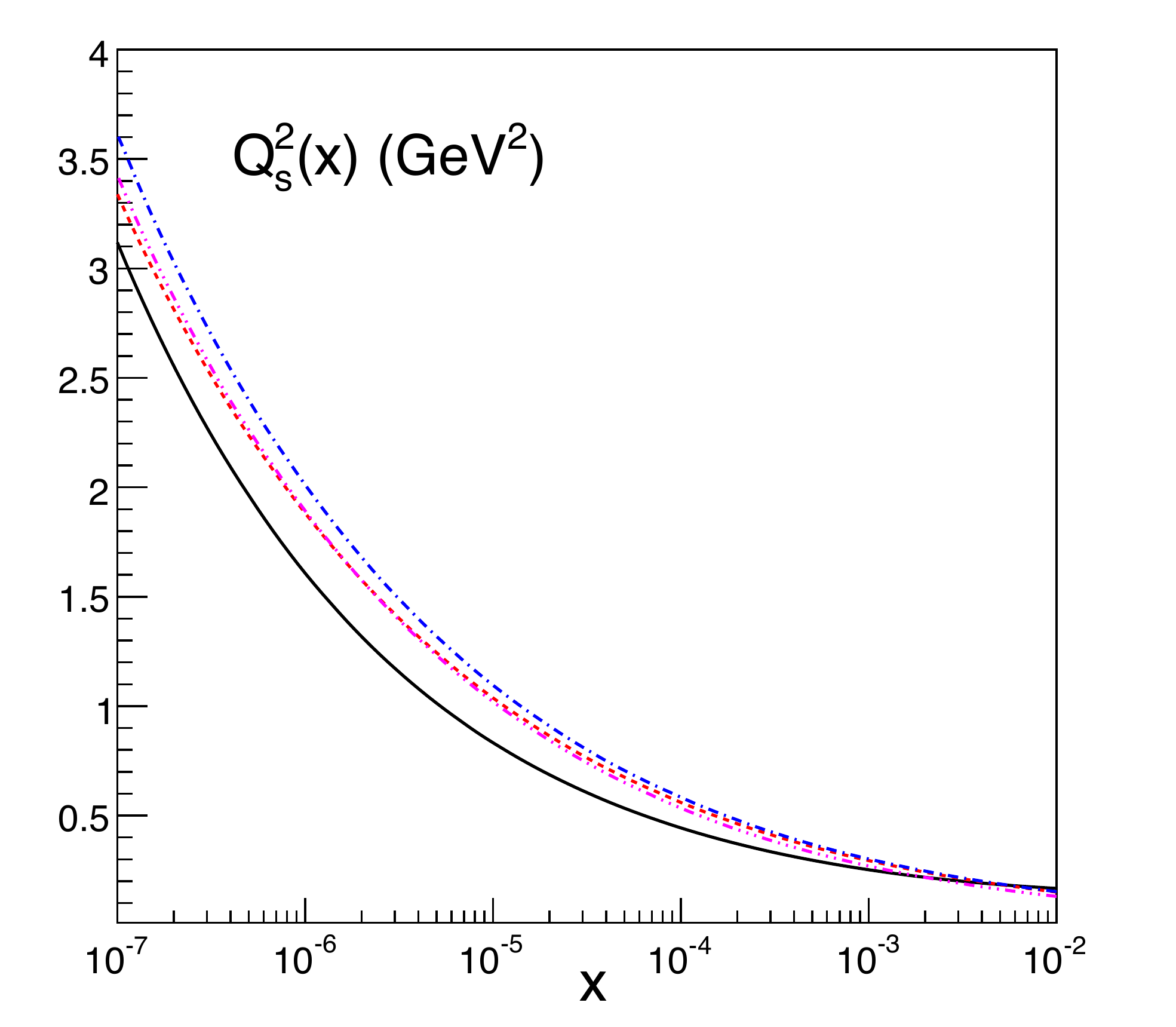}
\end{center}
\vspace*{-0.5cm}
\caption[a]{}
\label{n-qs2}
\end{figure}

Before addressing specific aspects of the fit results, it should be noted that all successful fits (those with $\chi^2/\text{d.o.f.}\lesssim 1.25$) yield a very similar value of the dipole scattering amplitude in the region of Bjorken-$x$ and $Q^2$ constrained  by experimental data. This is not an unexpected result since, once the calculation framework is fixed --the dipole model in our case-- the simultaneous description of the $x$ and $Q^2$-dependence of data leaves little room for variations of the optimal dipole amplitude. Indeed, the only fit parameter that does not relate directly to the behaviour of the dipole amplitude, either of its initial condition or its evolution, is the normalisation parameter $\sigma_0$.  We observe that the variations of the dipole amplitude among different fits are compensated by changes of the same order in the corresponding values of the $\sigma_0$ parameter,  $\sim10\%$. 
The mentioned compatibility among different evolution schemes is clearly seen in \fig{n-qs2}, left plot, which shows the dipole scattering amplitude resulting from four different fits at the highest $x$-value included in the fitted data set as a function of the dipole transverse size. The four dipole amplitudes represented in that plot deviate little from each other in the region of large and moderate dipole sizes. Using the ballpark estimate that relates the photon virtuality and the dipole size as $Q\sim 2/r$, fits up to $Q_{max}^2=500$ GeV$^2$ (there is a single point in the fitting set with $Q^2>500$ GeV$^2$) constrain the region of dipole sizes $r\gtrsim 10^{-1}$ GeV$^{-1}$. It is precisely outside this region, i.e for $r\lesssim 10^{-1}$ GeV$^{-1}$, that differences between the different parametrisation of the dipole amplitude start growing above $10\%$ and higher.

\fig{n-qs2}, right plot, shows the value of the saturation scale, obtained according to the definition given in \eq{qsx}, as a function of the rapidity variable $Y\!=\!\ln(x_0/x)$ for four different fits. Again, and in agreement with the previous discussion on the similarity of the dipole amplitudes, we observe that different fitting schemes lead to a very similar value for the saturation scale. In particular they all yield an initial value $Q_s^2(x_0=0.01)\approx 0.2$ GeV$^2$. This value is in very good agreement with the one obtained in previous fit to HERA data. Differences in the rate of growth of $Q_s^2(x)$ with decreasing $x$ are larger than those of the the dipole amplitude itself, with the rcBK-Bal scheme yielding the slowest evolution. 

For reference we have first performed fits using the BK equations with only running coupling corrections. The fit parameters are shown in \tabl{t1}. 
In this case it was possible to obtain good fits using only Balitsky's scheme for the running of the coupling. The other two schemes, PD and SD, lead to a too fast evolution speed that can not be reconciled with data. The fits are stable when changing the maximum virtuality in the fitted data set from $50$ to $650$ GeV$^2$.  The best fits are obtained when $\Delta Y_0$ is fixed to 0 and no restriction is imposed on the parameter $\gamma$, which tends to acquire values $\sim 1.14\div1.15$. It turns out that these range of $\gamma$ values lead to a non positive definite fourier transform of the dipole amplitude and, hence, to an unphysical unintegrated gluon distribution. In order to guarantee the physicality of the related gluon distribution we impose the constraint $\gamma\!<\!1.125$. We checked that below this limiting value the fourier transform of the initial conditions \eq{ic} is positive definite. This feature is preserved by the evolution, as it tends to decrease the effective {\it anomalous dimension} of the solutions down to its asymptotic value $\gamma\sim 0.85$ \cite{Albacete:2004gw,Triantafyllopoulos:2002nz}. This additional constraint leads to slightly higher $\chi^2/\text{d.o.f}$, but still or order $\mathcal{O}(1)$ and stable against increasing $Q^2_{max}$. It should be noted that the fits presented here vary slightly with respect to analogous fits presented in \cite{Albacete:2010sy,Albacete:2009fh}. The  main reason for this change of the fits results is the improvement of the numerical accuracy of the code and, mostly, to a more refined search strategy in the fitting algorithm that allows to find a larger amount of local minima.

Similarly, a very good description of data can be obtained using the DLA-rcBK equation \eq{dla} for the evolution of the dipole amplitude. The fit parameters are shown in \tabl{t2}. In this case, Balitsky's scheme for the running coupling yields the worst fits to data, $\chi^2/\text{d.o.f}\ge 1.5$. In turn, it is possible to get very good, stable with increasing $Q_{max}^2$ fits with he other two running coupling schemes, PD and SD. 
This can be understood as due to the  lower evolution speed yielded obtained within Balitsky's prescription with respect to the PD and SD schemes. The evolution speed is further reduced when the additional phase space restrictions entailed by the collinear resummation of \eq{dla} are also incorporated in the evolution kernel. The most remarkable feature of these fits is that they require large pre-evolution periods:  $2.9\!< \Delta Y_0 <\! 9.1$. This brings the physical solutions deeper into the scaling regime, thus speeding up the evolution and compensating the speed reduction induced by DLA effects. The fudge factor $C$ takes values larger than for rcBK fits: $C\sim 2.5\div3.8$.  
We have checked that, as suggested in \cite{Iancu:2015joa}, the effects of the DLA resummation on the initial conditions can be reabsorbed in to a slight reshuffling of the fit parameters. Nonetheless, all the fits described here have been obtained including the resummation effects on the initial condition, i.e. applying \eq{a} to the initial conditions defined in Eqs. (\ref{ic}) - (\ref{ic2}). The particular choice for the scale setting the running of the coupling in Eqns. \ref{a}-\ref{kdla} did not have any significant impact on the fit results. It should also be noted that the very small values of the initial scale $Q_0$, only slightly larger than $\Lambda_{QCD}$, do not carry any special physical meaning. Rather, the physically meaning full dipole amplitudes and the corresponding saturation scales are very similar to those obtained with rcBK evolution only, as already discussed at the beginning of this section and illustrated in \fig{n-qs2}.

Finally, fits using the kinematically corrected BK equation KC-rcBK \eq{kc} also provide a very good description of the data. The fit parameters are shown in \tabl{t3}. The discussion of the results here runs close to the one for the DLA-rcBK fits. Again, the use Balitsky's scheme yields reasonable fits only for $Q_{max}^2\!=\!50$ GeV$^2$, turning unstable for larger photon virtualities. In turn the use of PD or SD schemes for the running coupling kernel results in good, stable fits. However, and contrary to the rcBK and DLA-rcBK approaches, the quality of the fits worsens slightly when $Q_{max}^2$ is increased, revealing some tension when the fits are pushed towards the collinear limit. Again, KC-rcBK fits require a considerable amount of pre-evolution before comparing to experimental data: $<3.2\!<\Delta Y_0<\!6.6$, while the values obtained for the fudge factor are very similar to those for DLA-rcBK fits: $C\sim 2.5\div3.7$.

Left: Dipole scattering amplitude versus dipole size at $x=0.01$ for four of the fit parametrisations. Right: Saturation scale corresponding to the dipole amplitudes represented in the right plot as a function of Bjorken-$x$

\section{Conclusions}
To conclude, in this work we have shown that the two collinearly improved BK equations presented in \cite{Iancu:2015vea,Beuf:2014uia} provide a good description of HERA data on the reduced cross section measured in electron-proton collisions in the small Bjorken-$x$ region, $x\le 0.01$, and for moderate to large virtualities, $Q^2\!<\! 650$ GeV$^2$. This has been possible through the introduction of a new set of {\it pre-asymptotic} initial conditions. This exercise alone does not suffice to distinguish which of the two approaches studied here is better suited to describe the high-energy evolution of color dipoles. Rather, additional theoretical effort is needed in order to determine the right running coupling evolution scheme  and also to correct the factorisation theorems for particle production to the same degree of accuracy. Also, the phenomenological analysis of more exclusive processes is needed in order to shed more light on the compatibility of the different resummation procedures with experimental data.
\label{conclusions}

\section*{Acknowledgements}
I thank N\'estor Armesto, Guillaume Beuf and Edmond Iancu for helpful discussions. This work has been supported by the Spanish MINECO (FPA2013-47836 and Ram\'on y Cajal contract) and by the {\it QCDense} Career Integration Grant of the FP7 Framework Program for Research of the European Commission, reference number CIG/631558.


\begin{mcitethebibliography}{10}

\bibitem{Beuf:2014uia}
G.~Beuf,  \mbox{}  \href{http://dx.doi.org/10.1103/PhysRevD.89.074039}{{\em
  Phys.Rev.} {\bf D89} (2014)no.~7 074039}
  [\href{http://arXiv.org/abs/1401.0313}{{\tt arXiv:1401.0313 [hep-ph]}}]\relax
\mciteBstWouldAddEndPuncttrue
\mciteSetBstMidEndSepPunct{\mcitedefaultmidpunct}
{\mcitedefaultendpunct}{\mcitedefaultseppunct}\relax
\EndOfBibitem
\bibitem{Iancu:2015vea}
E.~Iancu, J.~Madrigal, A.~Mueller, G.~Soyez and D.~Triantafyllopoulos,  \mbox{}
   \href{http://dx.doi.org/10.1016/j.physletb.2015.03.068}{{\em Phys.Lett.}
  {\bf B744} (2015) 293} [\href{http://arXiv.org/abs/1502.05642}{{\tt
  arXiv:1502.05642 [hep-ph]}}]\relax
\mciteBstWouldAddEndPuncttrue
\mciteSetBstMidEndSepPunct{\mcitedefaultmidpunct}
{\mcitedefaultendpunct}{\mcitedefaultseppunct}\relax
\EndOfBibitem
\bibitem{Kovchegov:1999yj}
Y.~V. Kovchegov,  \mbox{}  {\em Phys. Rev.} {\bf D60} (1999) 034008
  [\href{http://arXiv.org/abs/hep-ph/9901281}{{\tt
  arXiv:hep-ph/9901281}}]\relax
\mciteBstWouldAddEndPuncttrue
\mciteSetBstMidEndSepPunct{\mcitedefaultmidpunct}
{\mcitedefaultendpunct}{\mcitedefaultseppunct}\relax
\EndOfBibitem
\bibitem{Balitsky:1996ub}
I.~Balitsky,  \mbox{}  {\em Nucl. Phys.} {\bf B463} (1996) 99
  [\href{http://arXiv.org/abs/hep-ph/9509348}{{\tt
  arXiv:hep-ph/9509348}}]\relax
\mciteBstWouldAddEndPuncttrue
\mciteSetBstMidEndSepPunct{\mcitedefaultmidpunct}
{\mcitedefaultendpunct}{\mcitedefaultseppunct}\relax
\EndOfBibitem
\bibitem{Jalilian-Marian:1997gr}
J.~Jalilian-Marian, A.~Kovner, A.~Leonidov and H.~Weigert,  \mbox{}  {\em Phys.
  Rev.} {\bf D59} (1998) 014014
  [\href{http://arXiv.org/abs/hep-ph/9706377}{{\tt
  arXiv:hep-ph/9706377}}]\relax
\mciteBstWouldAddEndPuncttrue
\mciteSetBstMidEndSepPunct{\mcitedefaultmidpunct}
{\mcitedefaultendpunct}{\mcitedefaultseppunct}\relax
\EndOfBibitem
\bibitem{Kovner:2000pt}
A.~Kovner, J.~G. Milhano and H.~Weigert,  \mbox{}  {\em Phys. Rev.} {\bf D62}
  (2000) 114005 [\href{http://arXiv.org/abs/hep-ph/0004014}{{\tt
  arXiv:hep-ph/0004014}}]\relax
\mciteBstWouldAddEndPuncttrue
\mciteSetBstMidEndSepPunct{\mcitedefaultmidpunct}
{\mcitedefaultendpunct}{\mcitedefaultseppunct}\relax
\EndOfBibitem
\bibitem{Weigert:2000gi}
H.~Weigert,  \mbox{}  {\em Nucl. Phys.} {\bf A703} (2002) 823
  [\href{http://arXiv.org/abs/hep-ph/0004044}{{\tt
  arXiv:hep-ph/0004044}}]\relax
\mciteBstWouldAddEndPuncttrue
\mciteSetBstMidEndSepPunct{\mcitedefaultmidpunct}
{\mcitedefaultendpunct}{\mcitedefaultseppunct}\relax
\EndOfBibitem
\bibitem{Kovchegov:2006vj}
Y.~Kovchegov and H.~Weigert,  \mbox{}  {\em Nucl. Phys. {\bf A}} {\bf 784}
  (2007) 188 [\href{http://arXiv.org/abs/hep-ph/0609090}{{\tt
  arXiv:hep-ph/0609090}}]\relax
\mciteBstWouldAddEndPuncttrue
\mciteSetBstMidEndSepPunct{\mcitedefaultmidpunct}
{\mcitedefaultendpunct}{\mcitedefaultseppunct}\relax
\EndOfBibitem
\bibitem{Balitsky:2006wa}
I.~I. Balitsky,  \mbox{}  {\em Phys. Rev. D} {\bf 75} (2007) 014001
  [\href{http://arXiv.org/abs/hep-ph/0609105}{{\tt
  arXiv:hep-ph/0609105}}]\relax
\mciteBstWouldAddEndPuncttrue
\mciteSetBstMidEndSepPunct{\mcitedefaultmidpunct}
{\mcitedefaultendpunct}{\mcitedefaultseppunct}\relax
\EndOfBibitem
\bibitem{Balitsky:2008zz}
I.~Balitsky and G.~A. Chirilli,  \mbox{}
  \href{http://dx.doi.org/10.1103/PhysRevD.77.014019}{{\em Phys. Rev.} {\bf
  D77} (2008) 014019} [\href{http://arXiv.org/abs/0710.4330}{{\tt
  arXiv:0710.4330 [hep-ph]}}]\relax
\mciteBstWouldAddEndPuncttrue
\mciteSetBstMidEndSepPunct{\mcitedefaultmidpunct}
{\mcitedefaultendpunct}{\mcitedefaultseppunct}\relax
\EndOfBibitem
\bibitem{Balitsky:2010ze}
I.~Balitsky and G.~A. Chirilli,  \mbox{}
  \href{http://dx.doi.org/10.1103/PhysRevD.83.031502}{{\em Phys.Rev.} {\bf D83}
  (2011) 031502} [\href{http://arXiv.org/abs/1009.4729}{{\tt arXiv:1009.4729
  [hep-ph]}}]\relax
\mciteBstWouldAddEndPuncttrue
\mciteSetBstMidEndSepPunct{\mcitedefaultmidpunct}
{\mcitedefaultendpunct}{\mcitedefaultseppunct}\relax
\EndOfBibitem
\bibitem{Beuf:2011xd}
G.~Beuf,  \mbox{}  \href{http://dx.doi.org/10.1103/PhysRevD.85.034039}{{\em
  Phys.Rev.} {\bf D85} (2012) 034039}
  [\href{http://arXiv.org/abs/1112.4501}{{\tt arXiv:1112.4501 [hep-ph]}}]\relax
\mciteBstWouldAddEndPuncttrue
\mciteSetBstMidEndSepPunct{\mcitedefaultmidpunct}
{\mcitedefaultendpunct}{\mcitedefaultseppunct}\relax
\EndOfBibitem
\bibitem{Chirilli:2011km}
G.~A. Chirilli, B.-W. Xiao and F.~Yuan,  \mbox{}
  \href{http://dx.doi.org/10.1103/PhysRevLett.108.122301}{{\em Phys.Rev.Lett.}
  {\bf 108} (2012) 122301} [\href{http://arXiv.org/abs/1112.1061}{{\tt
  arXiv:1112.1061 [hep-ph]}}]\relax
\mciteBstWouldAddEndPuncttrue
\mciteSetBstMidEndSepPunct{\mcitedefaultmidpunct}
{\mcitedefaultendpunct}{\mcitedefaultseppunct}\relax
\EndOfBibitem
\bibitem{Chirilli:2012jd}
G.~A. Chirilli, B.-W. Xiao and F.~Yuan,  \mbox{}
  \href{http://dx.doi.org/10.1103/PhysRevD.86.054005}{{\em Phys.Rev.} {\bf D86}
  (2012) 054005} [\href{http://arXiv.org/abs/1203.6139}{{\tt arXiv:1203.6139
  [hep-ph]}}]\relax
\mciteBstWouldAddEndPuncttrue
\mciteSetBstMidEndSepPunct{\mcitedefaultmidpunct}
{\mcitedefaultendpunct}{\mcitedefaultseppunct}\relax
\EndOfBibitem
\bibitem{Chirilli:2013rta}
G.~Chirilli, B.-W. Xiao and F.~Yuan,  \mbox{}
  \href{http://dx.doi.org/10.1016/j.nuclphysa.2013.02.147}{{\em Nucl.Phys.}
  {\bf A904-905} (2013) 841c}\relax
\mciteBstWouldAddEndPuncttrue
\mciteSetBstMidEndSepPunct{\mcitedefaultmidpunct}
{\mcitedefaultendpunct}{\mcitedefaultseppunct}\relax
\EndOfBibitem
\bibitem{Altinoluk:2014eka}
T.~Altinoluk, N.~Armesto, G.~Beuf, A.~Kovner and M.~Lublinsky,  \mbox{}
  \href{http://dx.doi.org/10.1103/PhysRevD.91.094016}{{\em Phys. Rev.} {\bf
  D91} (2015)no.~9 094016} [\href{http://arXiv.org/abs/1411.2869}{{\tt
  arXiv:1411.2869 [hep-ph]}}]\relax
\mciteBstWouldAddEndPuncttrue
\mciteSetBstMidEndSepPunct{\mcitedefaultmidpunct}
{\mcitedefaultendpunct}{\mcitedefaultseppunct}\relax
\EndOfBibitem
\bibitem{Altinoluk:2011qy}
T.~Altinoluk and A.~Kovner,  \mbox{}
  \href{http://dx.doi.org/10.1103/PhysRevD.83.105004}{{\em Phys.Rev.} {\bf D83}
  (2011) 105004} [\href{http://arXiv.org/abs/1102.5327}{{\tt arXiv:1102.5327
  [hep-ph]}}]\relax
\mciteBstWouldAddEndPuncttrue
\mciteSetBstMidEndSepPunct{\mcitedefaultmidpunct}
{\mcitedefaultendpunct}{\mcitedefaultseppunct}\relax
\EndOfBibitem
\bibitem{Albacete:2012xq}
J.~L. Albacete, A.~Dumitru, H.~Fujii and Y.~Nara,  \mbox{}
  \href{http://dx.doi.org/10.1016/j.nuclphysa.2012.09.012}{{\em Nucl.Phys.}
  {\bf A897} (2013) 1} [\href{http://arXiv.org/abs/1209.2001}{{\tt
  arXiv:1209.2001 [hep-ph]}}]\relax
\mciteBstWouldAddEndPuncttrue
\mciteSetBstMidEndSepPunct{\mcitedefaultmidpunct}
{\mcitedefaultendpunct}{\mcitedefaultseppunct}\relax
\EndOfBibitem
\bibitem{Stasto:2013cha}
A.~M. Stasto, B.-W. Xiao and D.~Zaslavsky,  \mbox{}
  \href{http://dx.doi.org/10.1103/PhysRevLett.112.012302}{{\em Phys.Rev.Lett.}
  {\bf 112} (2014)no.~1 012302} [\href{http://arXiv.org/abs/1307.4057}{{\tt
  arXiv:1307.4057 [hep-ph]}}]\relax
\mciteBstWouldAddEndPuncttrue
\mciteSetBstMidEndSepPunct{\mcitedefaultmidpunct}
{\mcitedefaultendpunct}{\mcitedefaultseppunct}\relax
\EndOfBibitem
\bibitem{Lappi:2015fma}
T.~Lappi and H.~M{\"a}ntysaari,  \mbox{}
  \href{http://dx.doi.org/10.1103/PhysRevD.91.074016}{{\em Phys.Rev.} {\bf D91}
  (2015)no.~7 074016} [\href{http://arXiv.org/abs/1502.02400}{{\tt
  arXiv:1502.02400 [hep-ph]}}]\relax
\mciteBstWouldAddEndPuncttrue
\mciteSetBstMidEndSepPunct{\mcitedefaultmidpunct}
{\mcitedefaultendpunct}{\mcitedefaultseppunct}\relax
\EndOfBibitem
\bibitem{Fadin:1995xg}
V.~S. Fadin, M.~I. Kotsky and R.~Fiore,  \mbox{}  {\em Phys. Lett.} {\bf B359}
  (1995) 181\relax
\mciteBstWouldAddEndPuncttrue
\mciteSetBstMidEndSepPunct{\mcitedefaultmidpunct}
{\mcitedefaultendpunct}{\mcitedefaultseppunct}\relax
\EndOfBibitem
\bibitem{Fadin:1998py}
V.~S. Fadin and L.~N. Lipatov,  \mbox{}  {\em Phys. Lett.} {\bf B429} (1998)
  127 [\href{http://arXiv.org/abs/hep-ph/9802290}{{\tt
  arXiv:hep-ph/9802290}}]\relax
\mciteBstWouldAddEndPuncttrue
\mciteSetBstMidEndSepPunct{\mcitedefaultmidpunct}
{\mcitedefaultendpunct}{\mcitedefaultseppunct}\relax
\EndOfBibitem
\bibitem{Ciafaloni:1998gs}
M.~Ciafaloni and G.~Camici,  \mbox{}  {\em Phys. Lett.} {\bf B430} (1998) 349
  [\href{http://arXiv.org/abs/hep-ph/9803389}{{\tt
  arXiv:hep-ph/9803389}}]\relax
\mciteBstWouldAddEndPuncttrue
\mciteSetBstMidEndSepPunct{\mcitedefaultmidpunct}
{\mcitedefaultendpunct}{\mcitedefaultseppunct}\relax
\EndOfBibitem
\bibitem{Aaron:2009wt}
{\bf H1} collaboration, F.~D. Aaron {\em et.~al.},  \mbox{}
  \href{http://dx.doi.org/10.1007/JHEP01(2010)109}{{\em JHEP} {\bf 01} (2010)
  109} [\href{http://arXiv.org/abs/0911.0884}{{\tt arXiv:0911.0884
  [hep-ex]}}]\relax
\mciteBstWouldAddEndPuncttrue
\mciteSetBstMidEndSepPunct{\mcitedefaultmidpunct}
{\mcitedefaultendpunct}{\mcitedefaultseppunct}\relax
\EndOfBibitem
\bibitem{Albacete:2009fh}
J.~L. Albacete, N.~Armesto, J.~G. Milhano and C.~A. Salgado,  \mbox{}
  \href{http://dx.doi.org/10.1103/PhysRevD.80.034031}{{\em Phys. Rev.} {\bf
  D80} (2009) 034031} [\href{http://arXiv.org/abs/0902.1112}{{\tt
  arXiv:0902.1112 [hep-ph]}}]\relax
\mciteBstWouldAddEndPuncttrue
\mciteSetBstMidEndSepPunct{\mcitedefaultmidpunct}
{\mcitedefaultendpunct}{\mcitedefaultseppunct}\relax
\EndOfBibitem
\bibitem{Albacete:2010sy}
J.~L. Albacete, N.~Armesto, J.~G. Milhano, P.~Quiroga~Arias and C.~A. Salgado,
  \mbox{}  \href{http://dx.doi.org/10.1140/epjc/s10052-011-1705-3}{{\em
  Eur.Phys.J.} {\bf C71} (2011) 1705}
  [\href{http://arXiv.org/abs/1012.4408}{{\tt arXiv:1012.4408 [hep-ph]}}]\relax
\mciteBstWouldAddEndPuncttrue
\mciteSetBstMidEndSepPunct{\mcitedefaultmidpunct}
{\mcitedefaultendpunct}{\mcitedefaultseppunct}\relax
\EndOfBibitem
\bibitem{Albacete:2012rx}
J.~Albacete, J.~Milhano, P.~Quiroga-Arias and J.~Rojo,  \mbox{}
  \href{http://dx.doi.org/10.1140/epjc/s10052-012-2131-x}{{\em Eur.Phys.J.}
  {\bf C72} (2012) 2131} [\href{http://arXiv.org/abs/1203.1043}{{\tt
  arXiv:1203.1043 [hep-ph]}}]\relax
\mciteBstWouldAddEndPuncttrue
\mciteSetBstMidEndSepPunct{\mcitedefaultmidpunct}
{\mcitedefaultendpunct}{\mcitedefaultseppunct}\relax
\EndOfBibitem
\bibitem{Kuokkanen:2011je}
J.~Kuokkanen, K.~Rummukainen and H.~Weigert,  \mbox{}  {\em Nucl.Phys.} {\bf
  A875} (2012) 29 [\href{http://arXiv.org/abs/1108.1867}{{\tt arXiv:1108.1867
  [hep-ph]}}]\relax
\mciteBstWouldAddEndPuncttrue
\mciteSetBstMidEndSepPunct{\mcitedefaultmidpunct}
{\mcitedefaultendpunct}{\mcitedefaultseppunct}\relax
\EndOfBibitem
\bibitem{Lappi:2013zma}
T.~Lappi and H.~M{\"a}ntysaari,  \mbox{}
  \href{http://dx.doi.org/10.1103/PhysRevD.88.114020}{{\em Phys.Rev.} {\bf D88}
  (2013) 114020} [\href{http://arXiv.org/abs/1309.6963}{{\tt arXiv:1309.6963
  [hep-ph]}}]\relax
\mciteBstWouldAddEndPuncttrue
\mciteSetBstMidEndSepPunct{\mcitedefaultmidpunct}
{\mcitedefaultendpunct}{\mcitedefaultseppunct}\relax
\EndOfBibitem
\bibitem{Albacete:2014fwa}
J.~L. Albacete and C.~Marquet,  \mbox{}
  \href{http://dx.doi.org/10.1016/j.ppnp.2014.01.004}{{\em
  Prog.Part.Nucl.Phys.} {\bf 76} (2014) 1}
  [\href{http://arXiv.org/abs/1401.4866}{{\tt arXiv:1401.4866 [hep-ph]}}]\relax
\mciteBstWouldAddEndPuncttrue
\mciteSetBstMidEndSepPunct{\mcitedefaultmidpunct}
{\mcitedefaultendpunct}{\mcitedefaultseppunct}\relax
\EndOfBibitem
\bibitem{Iancu:2015joa}
E.~Iancu, J.~D. Madrigal, A.~H. Mueller, G.~Soyez and D.~N. Triantafyllopoulos,
   \mbox{}  \href{http://arXiv.org/abs/1507.03651}{{\tt arXiv:1507.03651
  [hep-ph]}}\relax
\mciteBstWouldAddEndPuncttrue
\mciteSetBstMidEndSepPunct{\mcitedefaultmidpunct}
{\mcitedefaultendpunct}{\mcitedefaultseppunct}\relax
\EndOfBibitem
\bibitem{Albacete:2015zra}
J.~L. Albacete, J.~I. Illana and A.~Soto-Ontoso,  \mbox{}
  \href{http://arXiv.org/abs/1505.06583}{{\tt arXiv:1505.06583 [hep-ph]}}\relax
\mciteBstWouldAddEndPuncttrue
\mciteSetBstMidEndSepPunct{\mcitedefaultmidpunct}
{\mcitedefaultendpunct}{\mcitedefaultseppunct}\relax
\EndOfBibitem
\bibitem{Albacete:2007yr}
J.~L. Albacete and Y.~V. Kovchegov,  \mbox{}  {\em Phys. Rev.} {\bf D75} (2007)
  125021 [\href{http://arXiv.org/abs/arXiv:0704.0612 [hep-ph]}{{\tt
  arXiv:arXiv:0704.0612 [hep-ph]}}]\relax
\mciteBstWouldAddEndPuncttrue
\mciteSetBstMidEndSepPunct{\mcitedefaultmidpunct}
{\mcitedefaultendpunct}{\mcitedefaultseppunct}\relax
\EndOfBibitem
\bibitem{Gotsman:2004xb}
E.~Gotsman, E.~Levin, U.~Maor and E.~Naftali,  \mbox{}
  \href{http://dx.doi.org/10.1016/j.nuclphysa.2004.12.073}{{\em Nucl. Phys.}
  {\bf A750} (2005) 391} [\href{http://arXiv.org/abs/hep-ph/0411242}{{\tt
  arXiv:hep-ph/0411242 [hep-ph]}}]\relax
\mciteBstWouldAddEndPuncttrue
\mciteSetBstMidEndSepPunct{\mcitedefaultmidpunct}
{\mcitedefaultendpunct}{\mcitedefaultseppunct}\relax
\EndOfBibitem
\bibitem{Nikolaev:1990ja}
N.~N. Nikolaev and B.~G. Zakharov,  \mbox{}
  \href{http://dx.doi.org/10.1007/BF01483577}{{\em Z. Phys.} {\bf C49} (1991)
  607}\relax
\mciteBstWouldAddEndPuncttrue
\mciteSetBstMidEndSepPunct{\mcitedefaultmidpunct}
{\mcitedefaultendpunct}{\mcitedefaultseppunct}\relax
\EndOfBibitem
\bibitem{Mueller:1989st}
A.~H. Mueller,  \mbox{}  {\em Nucl. Phys.} {\bf B335} (1990) 115\relax
\mciteBstWouldAddEndPuncttrue
\mciteSetBstMidEndSepPunct{\mcitedefaultmidpunct}
{\mcitedefaultendpunct}{\mcitedefaultseppunct}\relax
\EndOfBibitem
\bibitem{Golec-Biernat:1998js}
K.~Golec-Biernat and M.~{W{\"u}sthoff},  \mbox{}  {\em Phys. Rev.} {\bf D59}
  (1998) 014017 [\href{http://arXiv.org/abs/hep-ph/9807513}{{\tt
  arXiv:hep-ph/9807513}}]\relax
\mciteBstWouldAddEndPuncttrue
\mciteSetBstMidEndSepPunct{\mcitedefaultmidpunct}
{\mcitedefaultendpunct}{\mcitedefaultseppunct}\relax
\EndOfBibitem
\bibitem{Albacete:2004gw}
J.~L. Albacete, N.~Armesto, J.~G. Milhano, C.~A. Salgado and U.~A. Wiedemann,
  \mbox{}  {\em Phys. Rev.} {\bf D71} (2005) 014003
  [\href{http://arXiv.org/abs/hep-ph/0408216}{{\tt
  arXiv:hep-ph/0408216}}]\relax
\mciteBstWouldAddEndPuncttrue
\mciteSetBstMidEndSepPunct{\mcitedefaultmidpunct}
{\mcitedefaultendpunct}{\mcitedefaultseppunct}\relax
\EndOfBibitem
\bibitem{Triantafyllopoulos:2002nz}
D.~N. Triantafyllopoulos,  \mbox{}  {\em Nucl. Phys.} {\bf B648} (2003) 293
  [\href{http://arXiv.org/abs/hep-ph/0209121}{{\tt
  arXiv:hep-ph/0209121}}]\relax
\mciteBstWouldAddEndPuncttrue
\mciteSetBstMidEndSepPunct{\mcitedefaultmidpunct}
{\mcitedefaultendpunct}{\mcitedefaultseppunct}\relax
\EndOfBibitem
\end{mcitethebibliography}
\bibliographystyle{JHEP-2modM}

\end{document}